\documentclass[aps,pre,twocolumn]{revtex4}
\expandafter\let\csname equation*\endcsname\relax
\expandafter\let\csname endequation*\endcsname\relax
\usepackage[T1]{fontenc}
\usepackage[utf8]{inputenc}
\usepackage{amsmath}
\usepackage{ulem}
\usepackage{graphicx,bbm,amssymb,amsopn}

\newcommand{\be}{\begin{equation}}
\newcommand{\ee}{\end{equation}}
\newcommand{\bea}{\begin{eqnarray}}
\newcommand{\eea}{\end{eqnarray}}

\begin{document}

\title[Few Islands Approximation of Hamiltonian System with divided Phase Space]{Few Islands Approximation of Hamiltonian System with divided Phase Space}

\author{Leonid A. Bunimovich$^1$, Giulio Casati$^{2,4}$, Toma\v{z} Prosen$^3$ and Gregor Vidmar$^3$}
\address{$^1$Georgia Institute of Technology, Atlanta, USA}
\address{$^2$Center for Nonlinear and complex systems, Universit\`a dell'Insubria, Como, Italy}
\address{$^3$Faculty of Mathematics and Physics, University of Ljubljana, Slovenia}
\address{$^4$International Institute of Physics, Federal University of Rio Grande do Norte, Natal, Brasil}

\begin{abstract}
It is well known that typical Hamiltonian systems have divided phase space consisting of regions with regular dynamics on KAM tori and region(s) with chaotic dynamics called chaotic sea(s). 
This complex structure makes rigorous analysis of such systems virtually impossible and significantly complicates numerical exploration of their dynamical properties. 
Hamiltonian systems with sharply divided phase space between regions of regular and chaotic dynamics are much easier to analyse, but there are only few cases or families of such systems known to date.
In this paper we outline a new approach for a systematic construction, starting from a generic KAM Hamiltonian system, of a system with a sharply divided phase space with an arbitrary number
of regular islands which are in one-to-one correspondence with islands of the initial KAM system. 
In this procedure a typical Hamiltonian system, for example a KAM billiard, is replaced by a sequence of Hamiltonian systems having an increasing (but finite) number of islands of regular motion. 
The islands in the substituting systems are sub-islands of the KAM islands in the initial system.
We apply this idea to two-dimensional lemon-shaped billiards, where the substituting systems are obtained by replacing parts of the curved boundaries by chords, so that in the limit of infinite number of islands the boundary of the
substituting system becomes arbitrary close to the original billiard's boundary.
\end{abstract}

\maketitle

\section{Introduction}

Our understanding of the dynamics of Hamiltonian systems with divided phase space is very limited thanks to the lacking of examples where a boundary between KAM regions and chaotic seas can be exactly determined. The only known exceptions are mushroom billiards and some piecewise linear discontinuous systems \cite{Bunimowich-01, Malovrh-02}. Therefore it is virtually impossible to understand how coexistence of various KAM islands influence the dynamics in chaotic region(s) (see e.g. \cite{Altmann-06,Ketzmerick,Robnik} for some representatives of the vast literature on the numerical studies of the problem). In this paper we develop an approach which allows to approximate generic billiards with coexistence of KAM tori and the chaotic sea by billiards with a finite number of KAM islands which are in strict correspondence with the islands of the original system. The sequence of billiards with the sharply divided phase-space approximates the original billiard in the sense that its boundary point-wise converges to the boundary of the original billiards.

Each sharp-boundary island in the approximating system is a sub-island of the one in the initial system, and typically has a substantially smaller phase-space area. 
We demonstrate that by considering approximating systems with increasing number of islands, 
systems with presumably sharply divided phase space are obtained, which contain more and more islands of the initial system which has an infinite number of KAM islands.

It is well known that a typical Hamiltonian system exhibits a mixed behavior, i.e. in some parts of its phase space the dynamics is regular while on the complementary part it is chaotic \cite{Lichtenberg-92}. This picture which everybody believes, even in the mathematical community, is however not rigorously proven to hold. The major obstacles are that totally different methods were developed to analyse regular (integrable) dynamics and chaotic dynamics and that it is virtually impossible to find a boundary between the regions with chaotic and those with regular dynamics. In fact, such boundary exists only in (i) 2-dimensional systems where a one dimensional torus separates the two-dimensional phase space into the interior and the exterior of the torus and (ii) in time-periodically perturbed one-dimensional systems possessing the separatrix, provided the perturbation is asymptotically weak and being resonant to the eigenoscillations at the boundary of the separatrix layer \cite{Soskin-09}. In higher dimensions, the KAM torus does not separate the phase space and completely different phenomena occur.
Even the construction of special concrete examples where the boundaries of KAM islands were exactly found and the dynamics rigorously analysed, was not very successful
with the exception of a very narrow class of high-dimensional mushroom billiards \cite{Bunimowich-06, Bunimowich-08}.

To be more precise, mushroom billiards \cite{Bunimowich-01} as well as
some piecewise linear symplectic maps \cite{Malovrh-02}, are the only examples which allowed to better understand some features of the dynamics in Hamiltonian systems with divided phase space. Classical \cite{Altmann-05, Tanaka-06} and  quantum  \cite {Dietz-06, Barnett-07, Dietz-07, Vidmar-07, Baecker-08} mushroom billiards were extensively studied both theoretically and experimentally.  However, these systems have a serious restriction which does not allow them to be considered as true representatives of Hamiltonian systems with divided phase space. Indeed, each KAM island sits in a specially constructed region in the phase space (a cap containing this island). Moreover, in typical Hamiltonian systems large islands correspond to lower order resonances and they are surrounded by chains of islands corresponding to higher order resonances. In mushroom billiards there are no such natural chains of islands.

In this paper we present a totally new approach for the construction of Hamiltonian systems with a finite number of KAM islands. This approach allows to approximate a given Hamiltonian system with divided phase space and with an infinite number of families of KAM tori (KAM islands in 2D) by a sequence of Hamiltonian systems with increasing (finite) number of KAM tori families (islands). We should note immediately (as already mentioned above) that the approximation of the original system has to be understood in a very weak sense (e.g. point-wise convergence of the billiard's boundaries) which does not imply approximation of the detailed structure of phase space.

The most important features of our approach are: (1) each island in each approximating system corresponds to an island in the initial Hamiltonian system (which has infinitely many islands) and (2) all the approximating systems  presumably have sharply divided phase space. By making accurate numerical analysis we demonstrate that each island in the approximating systems is included in the corresponding island of the initial system.
The main idea of our procedure is exactly based on the fact that in typical Hamiltonian systems each family of KAM tori is surrounded by other KAM families corresponding to higher resonances.
Therefore, by cutting the boundary around 
stable periodic orbit which gives rise to the main island we destroy all satellite islands (resonances) and only the central (main) island remain.

We demonstrate the efficiency of our approach on the example of lemon-shaped billiards. Besides being the most paradigmatic and popular models of generic Hamiltonian dynamical systems, the billiards allow to operate only in configuration space in order to obtain the appropriate sequence of approximating billiards with finite number of islands.

\section{Lemon Billiards}

\begin{figure}
\centering	
\includegraphics[width=\columnwidth]{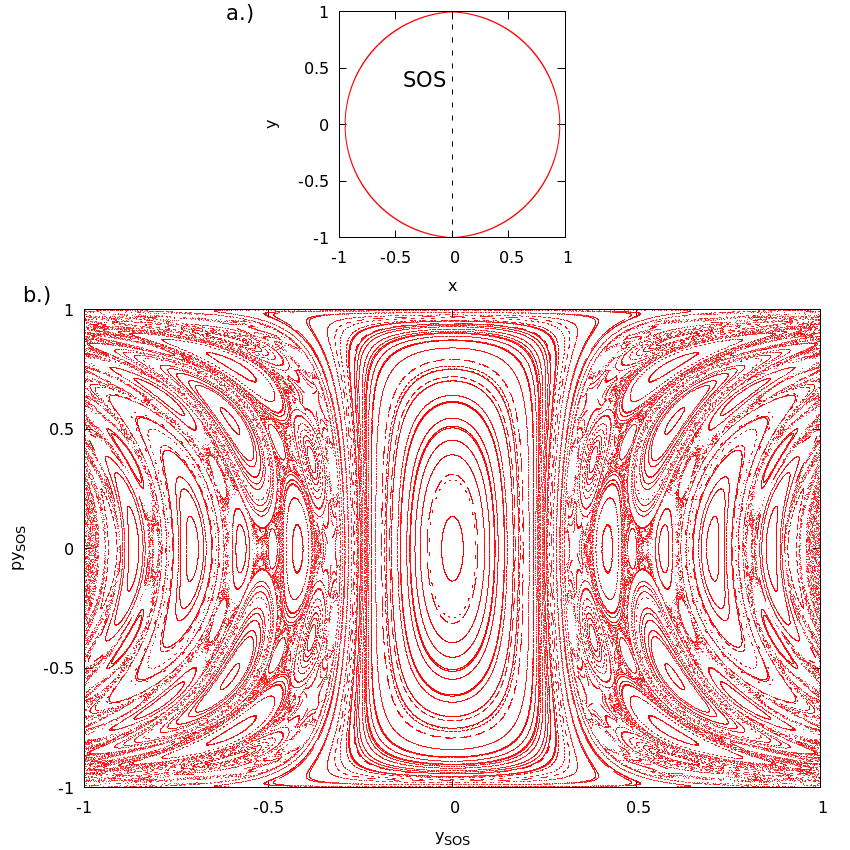}
\vspace{-1mm}
\caption{Lemon billiard: a) configuration space (x, y) with $0 < x < 0.95 = (1 - B)$ and $0 < y < 0.999$ ($R=1$);
b) SOS: $y_{SOS}$ (abscisa), $(p_y)_{SOS}$ (ordinate) at each bounce at $x = 0$.}
\label{Figure1-limonca}
\end{figure}

Our starting system is a lemon billiard (Fig.\ref{Figure1-limonca}) whose classical and quantum dynamics has been extensively studied \cite {Heller-93, Lopac-99, Lopac-01, Makino-01, Chen-13,
Bunimowich-15}. A lemon billiard table results from intersecting two identical circles with radius $R$ and with the distance $2B$ between their centers being less than the diameter of the circles. The boundary of a lemon billiard is described by the following implicit equations:

\begin{align}
(x + B)^2 + y^2 = R^2, \quad x \ge 0, \nonumber \\
(x - B)^2 + y^2 = R^2, \quad x < 0,
\end{align}

In Fig.\ref{Figure1-limonca} we show the configuration space and the surface of section (SOS) of our lemon billiard with $B = 0.05$ and $R = 1$. This is the only lemon billiard configuration which will be used in the present paper. The points on the Poincar\' e Surface of section (SOS) are plotted each time the billiard particle crosses the line $x = 0$. The SOS phase portrait of the lemon billiard has regular islands with fractal structure of the resonant islands around the main islands. This billiard is an example of a typical Hamiltonian system which contains an infinite number of KAM islands.
The islands correspond to periodic orbits $m:n$ of the (full lemon) billiard: $m$ counts the number of revolutions of the orbit around the center before closing on itself, $n$ counts the total number of bounces with the boundary.

Due to geometric reflection symmetry only one half of the billiard, $x\ge 0$, will be considered in the following.

\section{Cut-off billiards}
\label{Chap:Cut-off}

\begin{figure}
\centering	
\includegraphics[width=\columnwidth]{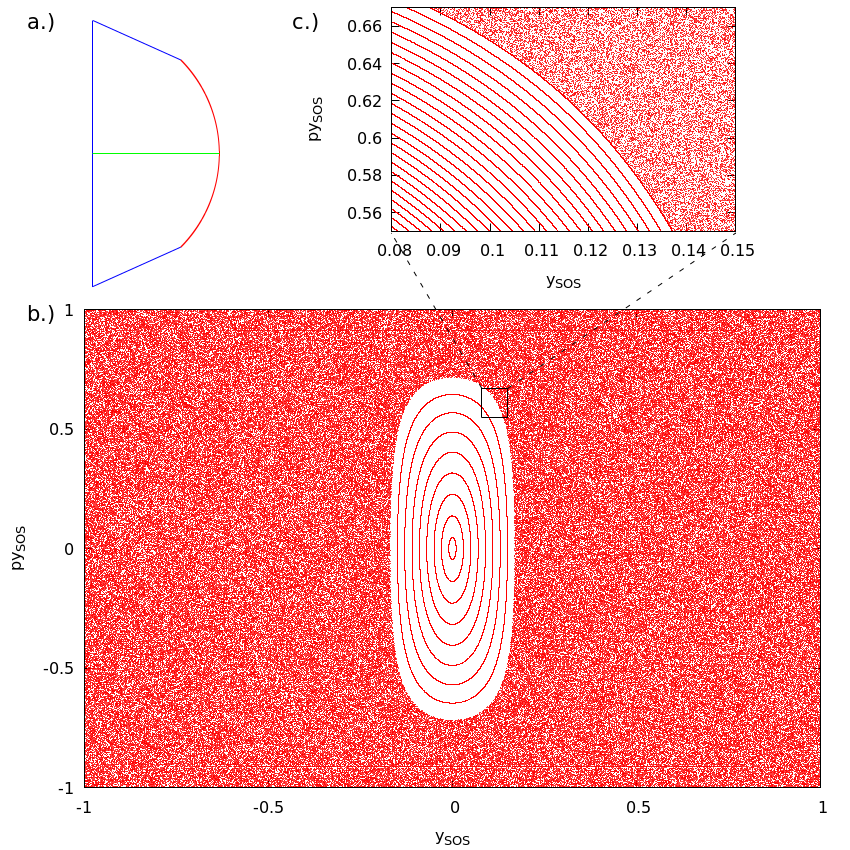}
\vspace{-1mm}
\caption{Cut-off billiard No. 1 - one periodic orbit included: a) billiard boundary with 1:2 periodic orbit shown, and $\epsilon_{1:2} = 0.78$;
green = periodic orbit, red = circular parts of the billiard's boundary, blue = chord parts of the billiard's boundary;
b) SOS at bounces at $x = 0$; c) enlargement at the border of the island of periodic orbit 1:2.
}
\label{Figure2-cutOffNo1}
\end{figure}

The sequence of so-called cut-off billiards (that may asymptotically approximate the shape of the original lemon billiard) is generated in the following way: the boundary of the lemon billiard with $B = 0.05$ and $R = 1$ is cut with several lines (chords) such that a chosen set of stable periodic orbits of the lemon billiard remain and all other stable periodic orbits of the lemon billiard are eliminated (by chords). The circular arcs -- parts of the original lemon boundary -- of  length $\epsilon_{m:n}$, symmetrically placed around the points where the periodic orbit $m:n$ touches the boundary of the lemon billiard, are connected with straight lines. Five such billiards have been constructed for our example and their SOSs calculated: they are labeled as cut-off billiard No.1, No.2 --- No.5,
where the cut-off billiard No.$k$ includes the first $k$ most relevant stable periodic orbits. If more (than one) periodic orbits touch the arc of the billiard at the same point, then $\epsilon_{m:n}$ with the lower $n$ is taken into account. In our calculations of the SOS we have considered with particular
attention the boundaries between the regular islands and the chaotic sea, and the possibility of secondary (and higher) resonant islands have been carefully investigated.

Fig. \ref{Figure2-cutOffNo1} shows the configuration space and the SOS at $x = 0$ of the cut-off billiard No. 1 with $\epsilon_{1:2} = 0.78$ which is the largest allowed value $\epsilon_{1:2} = \pi/4$ around the periodic orbit 1:2 with accuraccy to 2 decimal places, while at $\epsilon_{1:2} > \pi/4$ the periodic orbit 1:4 starts to appears. In Fig. \ref{Figure2-cutOffNo1} c) we present an enlarged view of the boundary between the regular island and chaotic sea. One can see that in the cut-off billiard No. 1 no secondary resonant island appears around the main island. We have performed even much more careful numerical checks of this observation than shown here, and up to machine resolution with double-precision floating point arithmetic, we have indeed found no traces of the resonant island chains.

\begin{figure}
\centering	
\includegraphics[width=\columnwidth]{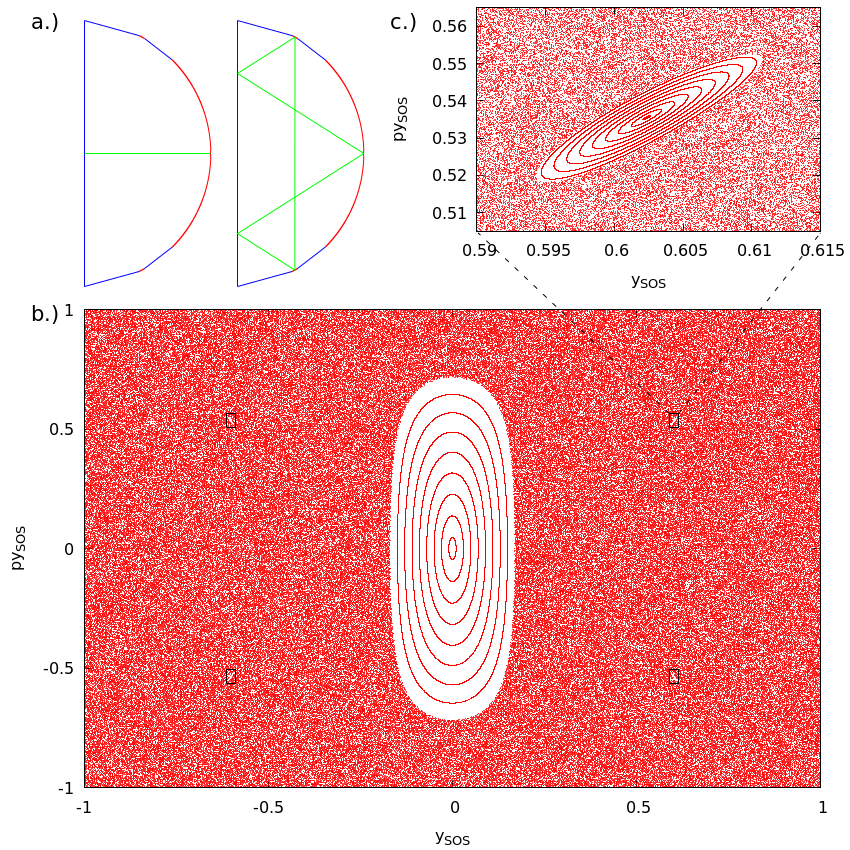}
\vspace{-1mm}
\caption{Cut-off billiard No. 2 - 2 periodic orbits included: a) billiard boundary with periodic orbits 1:2 and 1:3 shown, and $\epsilon_{1:2} = 0.78$,  $\epsilon_{1:3} = 0.020$;
green = periodic orbit, red = circle part of the billiard, blue = linear parts of the billiard's boundary;
b) SOS at bounces at $x = 0$: c) enlargement of one of the 1:3 islands and its boundary with the chaotic sea.
}
\label{Figure3-cutOffNo2}
\end{figure}

Fig. \ref{Figure3-cutOffNo2} shows the configuration space and the SOS of the phase space at $x = 0$ of the cut-off billiard No. 2 with $\epsilon_{1:2} = 0.78$ around the periodic orbit 1:2 and $\epsilon_{1:3} = 0.020$ around the periodic orbit 1:3. The reason for the choice of the $\epsilon_{1:2} = 0.78$ is the same as in cut-off billiard No 1. The $\epsilon_{1:3} = 0.020$ is the largest allowed $\epsilon_{1:3}$ with accuracy to 3 decimal places where the periodic orbit 1:6 does not appear yet. In Fig. \ref{Figure3-cutOffNo2} c) an enlargement of the boundary between a regular island from the periodic orbit 1:3 and  the chaotic sea is shown additionally. One sees that no secondary resonances appear. From the periodic orbit 1:3 there are 4 mirror regular islands sitting at the centers $y_0=\pm 0.6024$ and $(p_y)_0 = \pm 0.536$. In the same way the boundary between the central regular island from the periodic orbit 1:3 and the chaotic sea has been checked. Due to the symmetric mirror positions of other 3 regular 1:3 islands one can conclude that no secondary resonant islands around the periodic orbits' islands in the cut-off billiard No. 2 appear.

\begin{figure}
\centering	
\includegraphics[width=\columnwidth]{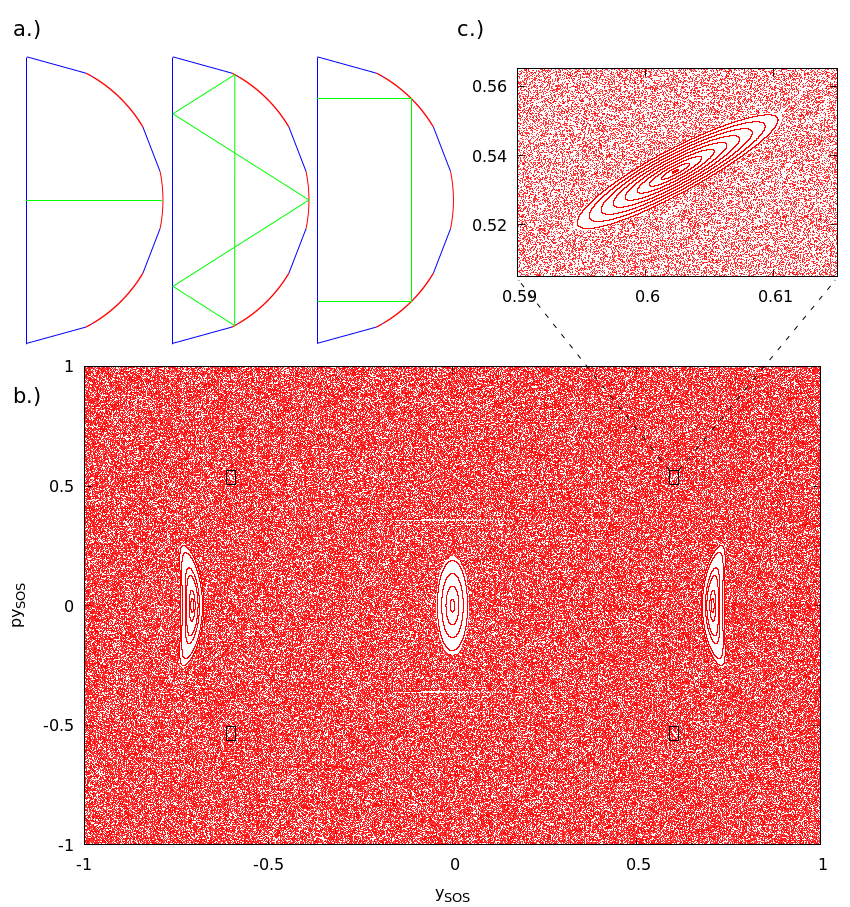}
\vspace{-1mm}
\caption{Cut-off billiard No. 3 - 3 periodic orbits included: a) billiard boundary with periodic orbits 1:2, 1:3 and 1:4 shown, and $\epsilon_{1:2} = 0.2$, $\epsilon_{1:3} = 0.020$,  $\epsilon_{1:4} = 0.25$; green = periodic orbit, red = circle part of the billiard, blue = linear parts of the billiard's boundary;
b) SOS at bounces at $x = 0$; c) enlargement of one of the 1:3 islands of stability and its boundary with the chaotic sea.
}
\label{Figure4-cutOffNo3}
\end{figure}

Fig.\ref{Figure4-cutOffNo3} shows the configuration space and the SOS at $x = 0$ of the cut-off billiard No. 3 with $\epsilon_{1:2} = 0.2$, $\epsilon_{1:3} = 0.020$ and $\epsilon_{1:4} = 0.25$. The reason for the choice of  $\epsilon_{1:3} = 0.020$ is the same as in the cut-off billiard No. 2. The largest possible $\epsilon_{1:4} = 0.25$ was chosen so that curved arcs from $\epsilon_{1:3}$ and $\epsilon_{1:4}$ do not overlap, whereas the choice of $\epsilon_{1:2} = 0.2$ is such that stickiness would qualitatively not appear. The stickiness which results from partial barriers to transport in phase space (such as cantori) is not a desired property here, because it would make it more difficult to detect secondary resonances around the main islands. Therefore a lower value than the maximally allowed one for the parameter
$\epsilon_{1:2}$ has been taken (the maximally allowed $\epsilon_{1:2}$ determined by the condition that the higher order islands do not appear, is $\epsilon_{1:2} = 0.379$). We will discuss more about stickiness in section \ref{Chap:stickiness}.
It should be stressed, however, that the phenomenon of stickiness does not in principle compromise the main idea of the present paper.
In Fig. \ref{Figure4-cutOffNo3} c)  an  enlargement of the boundary between (one of the 4) regular island(s) from the periodic orbit 1:3 and the chaotic sea is shown additionally. One sees again that around the enlarged island 1:3 no secondary resonant islands appear. In the same way the boundary between the central regular island from periodic orbit 1:2 and the chaotic sea has been checked as well the boundary between regular islands from the periodic orbit 1:4 (2 mirror islands centerd at $y_0=\pm \pi/4$ and $(p_y)_0 = 0$) and chaotic sea. From all our empirical data we can conclude that no secondary resonant islands  appear around the main periodic orbits' islands in cut-off billiard No. 3.

\begin{figure}
\centering	
\includegraphics[width=\columnwidth]{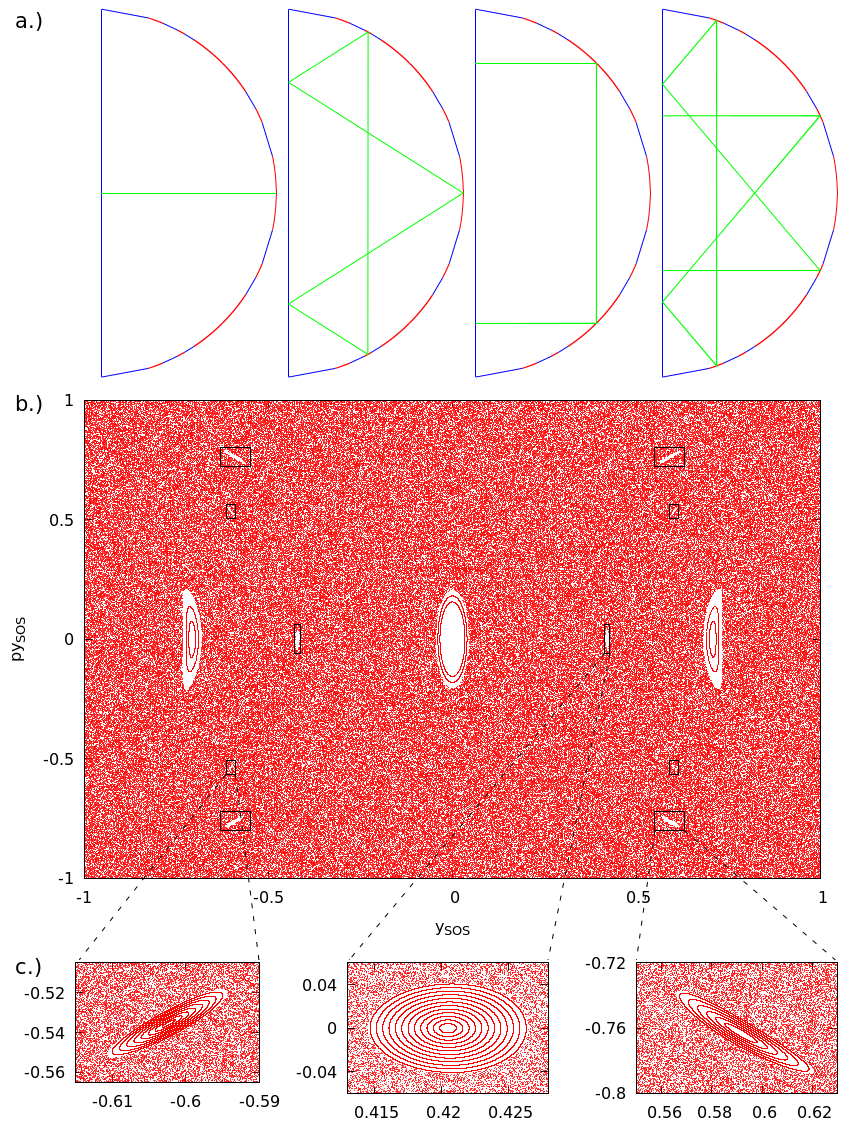}
\vspace{-1mm}
\caption{Cut-off billiard No. 4 - 4 periodic orbits included; a) billiard boundary with periodic orbits 1:2, 1:3, 1:4 and 3:8 shown, and where $\epsilon_{1:2} = 0.2$, $\epsilon_{1:3} = 0.020$, $\epsilon_{1:4} = 0.14$ and $\epsilon_{3:8} = 0.041$; green = periodic orbit, red = circle part of the billiard, blue = linear parts of the billiard's boundary; b) SOS at bounces at $x = 0$; c) enlargement of the 1:3 island (left) and two of the 3:8 islands (middle and right) of stability and their boundary with the chaotic sea.
}
\label{Figure5-cutOffNo4}
\end{figure}

\begin{figure}
\centering	
\vspace{-1mm}
\includegraphics[width=\columnwidth]{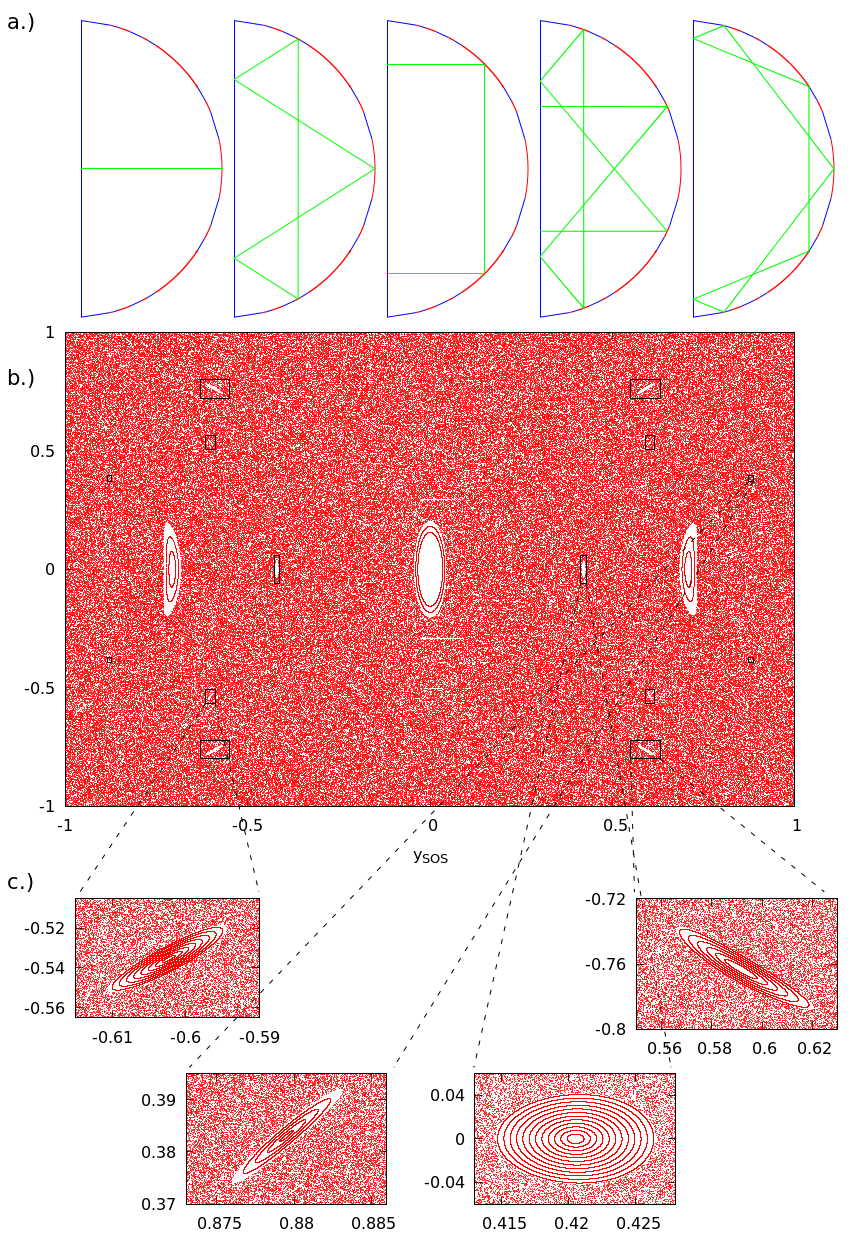}
\caption{Cut-off billiard No. 5 - 5 periodic orbits included; a) billiard boundary with periodic orbits 1:2, 1:3, 1:4, 3:8 and 1:5 shown, and where $\epsilon_{1:2} = 0.2$, $\epsilon_{1:3} = 0.020$, $\epsilon_{1:4} = 0.14$, $\epsilon_{3:8} = 0.041$  and $\epsilon_{1:5} = 0.010$; green = periodic orbit, red = circle part of the billiard, blue = linear parts of the billiard's boundary;
b) SOS at bounces at $x = 0$; c) enlargement of the 1:3 island (upper left), two of the 3:8 islands (bottom right and upper right) and 1:5 island (bottom left) and their boundary with the chaotic sea.
}
\label{Figure6-cutOffNo5}
\end{figure}

Fig. \ref{Figure5-cutOffNo4} shows the configuration space and the SOS at $x = 0$ of the cut-off billiard No. 4 with $\epsilon_{1:2} = 0.2$, $\epsilon_{1:3} = 0.020$, $\epsilon_{1:4} = 0.14$ and $\epsilon_{3:8} = 0.041$. The reason for the choice of the $\epsilon_{1:2} = 0.2$ and $\epsilon_{1:3} = 0.020$ is the same as in cut-off billiards No. 2 and No. 3. The reason for the choice of the $\epsilon_{1:4}=0.14$ is elimination of higher order islands' set associated to a higher order periodic orbit. The $\epsilon_{3:8} = 0.041$ is the largest possible value so that periodic orbit 1:8 and another set of islands associated to another higher order periodic orbit do not appear in the system.
In Fig. \ref{Figure5-cutOffNo4} c) enlargement of the boundary between (one of the 4) regular island(s) from the periodic orbit 1:3 and chaotic sea is shown additionally and enlargement of both kinds of islands from the periodic orbit 3:8 (6 island all together): 2 mirror islands with the centers at $y_0=\pm 0.4208$, $(p_y)_0 = 0$ and 4 mirror islands with the centers at $y_0=\pm 0.591$, $(p_y)_0 = \pm 0.761$. One again sees that around all the enlarged islands no secondary resonant islands   appear. In the same way the boundary between the central regular island from the periodic orbit 1:2 and chaotic sea as well as the boundary between regular islands from the periodic orbit 1:4 and chaotic sea have been checked. From all our data we can conclude that no secondary resonant islands appear around the main periodic orbits' islands in the cut-off billiard No. 4.

Fig. \ref{Figure6-cutOffNo5} shows the configuration space and the SOS at $x = 0$ of the cut-off billiard No. 5 with $\epsilon_{1:2} = 0.2$, $\epsilon_{1:3} = 0.020$, $\epsilon_{1:4} = 0.14$, $\epsilon_{3:8} = 0.041$ and $\epsilon_{1:5} = 0.010$. The reason for the choice of the $\epsilon_{1:2}, \epsilon_{1:3}, \epsilon_{1:4}$ and $\epsilon_{3:8}$ is the same as in cut-off billiard No. 4.
The reason for the choice of $\epsilon_{1:5} = 0.010$ is that it is the largest possible $\epsilon_{1:5}$ with accuracy to 3 decimal places so that periodic orbit 2:5 does not appear yet.
In Fig. \ref{Figure6-cutOffNo5} c) an enlargement of the boundary between small regular islands and chaotic sea is shown additionally like in in Fig. \ref{Figure5-cutOffNo4} c). The new islands from periodic orbit 1:5 have the centers at $y_0=\pm 0.879$, $(p_y)_0 = \pm 0.383$. One can see that in the cut-off billiard No. 5, around all the enlarged islands, no secondary resonant islands appear. In the same way the boundary between the central regular island from the periodic orbit 1:2 and chaotic sea as well as the boundary between regular islands from the periodic orbit 1:4 and chaotic sea have been checked. Again, from our data we can conclude that no secondary resonant islands appear around the periodic orbits' islands in cut-off billiard No. 5.


From the above procedure of the construction of the cut-off billiards No. 1, No. 2, No. 3, No. 4 and No. 5 it is obvious that it can only be done in such a way that each next cut-off billiard No.$(k+1)$ has, apart from the new regular island(s) that appeared, also smaller sized (or in some cases equal) corresponding islands in comparison to the previous step (in cut-of billiard No.$k$). The previous islands have to be made smaller in order to prevent emergence of higher order resonant islands (without a sharp boundary) as well as to avoid stickiness due to emergence of cantori (which is the precursor of higher order KAM island chains).
It is also clear that the length of the largest chord becomes smaller with increasing $k$ in this approximating billiard sequence, since we want more and more regular orbits to be included which populate the boundary densely in the limit. Based on reported numerical observations for the first five cut-off billiards, we conjecture that our conclusions (namely on the sharpness of the island's boundary) will apply also to higher order billiards No.$k$.

\section{Phase space portraits of the first chord-hit times}
\label{Chap:collisions}

We shall now attempt to quantitatively characterize the agreement in dynamics between the original KAM billiard and the set of cut-off billiards. For each trajectory, the agreement between the corresponding pair of billiard dynamics' is perfect as long as no chord is hit, namely as long as all collisions with the boundary happen on circular arcs. However, when the trajectory of the cut-off billiard hits the chord, it enters the chaotic domain and at the same time deviates from the trajectory of the original billiard. Nevertheless, the deviation is smaller the shorter the chord is -- since shorter chords are more tangential to the boundary -- and the maximal length of the chord decreases with increasing $k$ in the cut off billiard sequence. Therefore, one has better and better agreement of dynamics up to any fixed number of collisions with increasing $k$.

\begin{figure}
\includegraphics[width=\columnwidth]{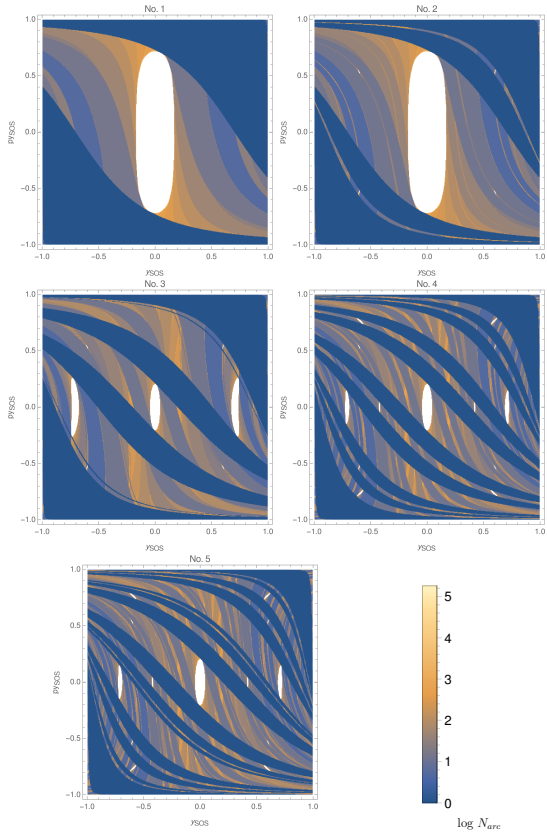}
\caption{Phase space (surface of section) color plots of the number of collisions (as a function of orbit's initial phase space coordinate) with the arcs before the first collision with any straight line segment (chord) -- i.e. the first chord-hit times, for all 5 consecutive cut-off billiards analysed in the paper. Resulting numbers are presented in logarithmic scale, see scale on the vertical bar. Note that the region of SOS where the "survival time" (time in number of collisions where the billiard motion follows the original KAM billiard) is smaller than some fixed number, seems to become a fractal set with zero measure.}
\label{Figure7-collisions-PS}
\end{figure}

We plot in Fig. \ref{Figure7-collisions-PS} the first chord-hit time SOS charts for all the five cut-off billiards that we have studied previously. The figure illustrates nicely how the sharp phase space boundary is formed 
between regions when the first chord-hit times are finite and infinite (white regions). Progressing in the cut off billiard sequence No. $k$, the set with first chord-hit times smaller than a fixed predetermined value $t^*$ seems to become fractal.

\section{Stickiness}
\label{Chap:stickiness}

Stickiness is a property of typical  dynamical systems where chaotic orbits are hindered by cantori to escape from the region close to a regular island \cite{Contopoulos-71}. It implies a temporary concentration of chaotic orbits in regions smaller than the entire asymptotically accessible chaotic region. The stickiness in mushroom billiards was studied in \cite{Altmann-05, Tanaka-06, Dettmann-11, Bunimowich-14}.

\begin{figure}
\centering	
\vspace{-1mm}
\includegraphics[width=\columnwidth]{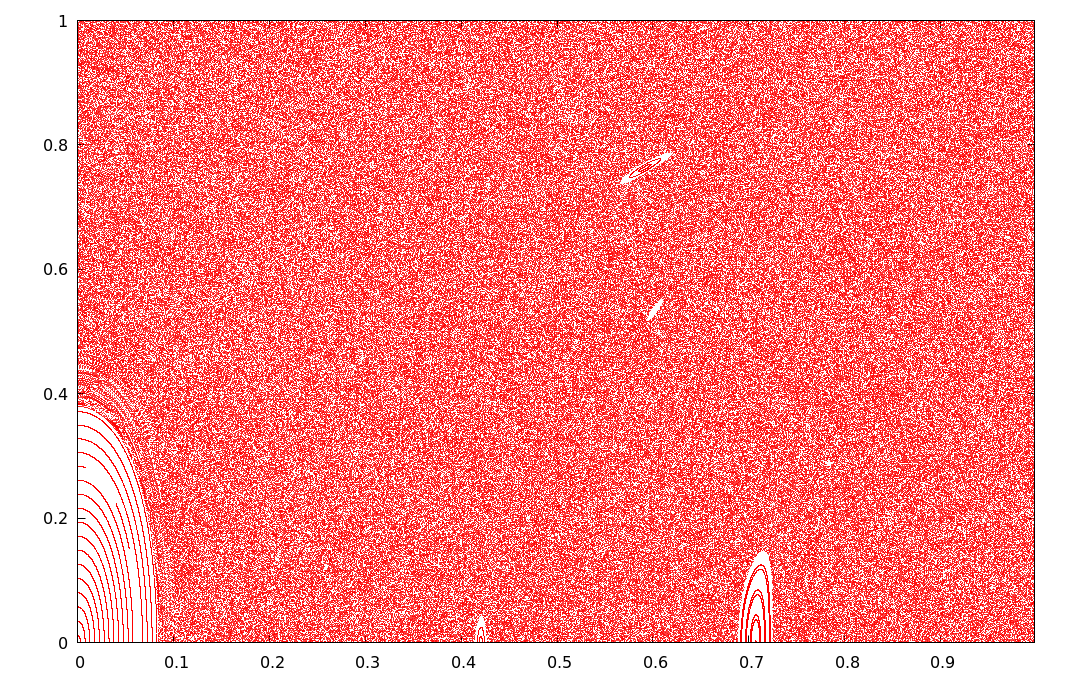}
\caption{Quarter of (de-symetrized) SOS of cut-off billiard No. 5, for a sticky case with $\epsilon_{1:2} = 0.379$, $\epsilon_{1:3} = 0.020$, $\epsilon_{1:4} = 0.14$, $\epsilon_{3:8} = 0.041$ and $\epsilon_{1:5} = 0.01$.
}
\label{Figure8-cutOffNo5-stickiness-PS}
\end{figure}

As mentioned above, if some curved boundary sections corresponding to appropriate periodic orbits in our cut-off billiards are too large, then stickiness appears (see Fig. \ref{Figure8-cutOffNo5-stickiness-PS} where $\epsilon_{1:2} = 0.379$).

\begin{figure}
\centering	
\vspace{-1mm}
\includegraphics[width=\columnwidth]{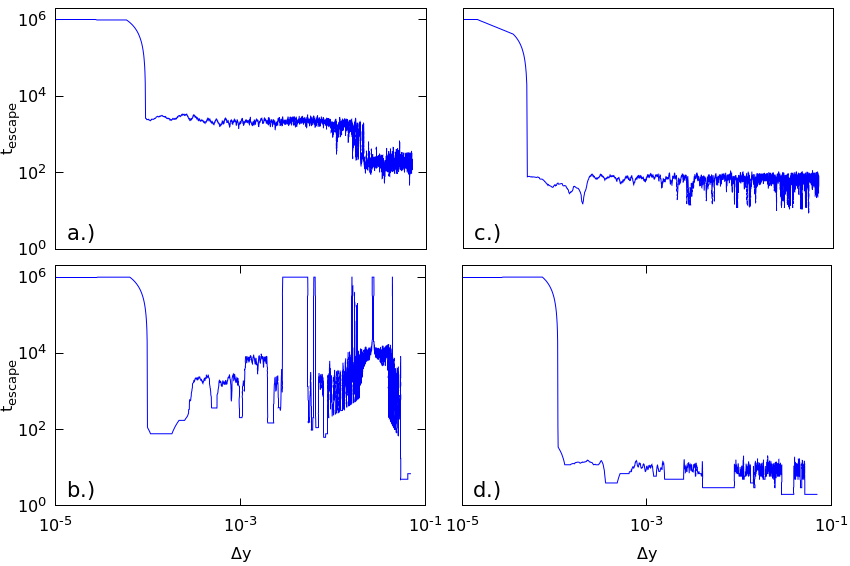}
\caption{Escape times $t_{\rm escape}$ as a function of the distance $\Delta y=y - y_a$ from an initial point $(y_a,p_y=0)$ placed near the edge of an island:
a) from the set $\{|y| < 0.45\} \bigcup \{ |p_y| > 0.5\}$
for cut-off billiard No. 5 from Fig. \ref{Figure8-cutOffNo5-stickiness-PS} around the 1:2 island with $y_a= 0.0841$;
b) from the set $\{|y| < 0.45\} \bigcup \{|p_y| > 0.5\}$ for the lemon billiard around the 1:2 island and with $y_a = 0.3042$;
c) from the set $\{|y| < 0.45\} \bigcup \{|p_y| > 0.5\}$ for cut-off billiard No. 5 from Fig. \ref{Figure6-cutOffNo5} around the 1:2 island with $y_a= 0.04462$;
d) from the set $\{|y| < 0.8\}$ for cut-off billiard No. 5 from Fig. \ref{Figure8-cutOffNo5-stickiness-PS} around the 1:4 island with $y_a= 0.72515$. In all plots we show moving averages over $50$ consecutive points which reduces oscillations and retains high resolution of the figures.
}
\label{Figure9-stickiness-escapes}
\end{figure}

We quantitatively describe the stickiness by calculating the time $t_{escape}$ an orbit in chaotic sea (close to the boundary of the regular island) needs to escape from a given region.
The results are shown in Fig. \ref{Figure9-stickiness-escapes} for four different cases. On the abscissa we plot, for varying $y$, the distance $\Delta y=y - y_a$ from the most left starting point $y_a$, which is slightly smaller than the regular island border $y^*$. On the ordinate we plot the escape time of the orbit with the initial condition $(y,p_ y=0)$ on SOS and escaping from the region $\{(y,p_y);|y| < y_{max} \bigcup |p_y| > (p_y)_{max}\}$. The region of $y$-coordinates of the initial conditions between $y=y_a$ and $y=y_b = y_a + 0.07$ has been divided into $10^5$ equidistant starting points. Each orbit runs maximally for $10^6$ iterations, i.e. returns to the SOS.

In Fig. \ref{Figure9-stickiness-escapes} a) escape times $t_{escape}$ around the periodic orbit 1:2 for the sticky case with 5 periodic orbits from Fig. \ref{Figure8-cutOffNo5-stickiness-PS} are shown. In Fig. \ref{Figure9-stickiness-escapes} c) the "non-sticky" case from No. 5 cut-off billiard from section \ref{Chap:Cut-off} (Fig. \ref{Figure6-cutOffNo5}) around the same periodic orbit 1:2 is shown. One can see that in the neighbourhood of a regular island the "non-sticky" cut-off billiard has more than an order of magnitude lower escape times in comparison to "sticky" one, which indeed proves the stickiness. In Fig. \ref{Figure9-stickiness-escapes} b) escape times for lemon billiard around periodic orbit 1:2 are presented. In the plot existences of few small higher order regular islands as well as stickiness are detected via orders of magnitude increases of escape times. In Fig. \ref{Figure9-stickiness-escapes} d) the same billiard as in Fig. \ref{Figure9-stickiness-escapes} a) is presented, but now the region around island 1:4 has been investigated - no stickiness was found in this last case.

Summarizing, our study indicates that for cuts with sufficiently long straight lines around the boundary-collision points corresponding to a set of pre-selected (say shortest) stable periodic orbits, one obtains a non-sticky chaotic component with only the desired finitely many KAM island chains. By decreasing the sizes of cuts we eventually obtain stickeness either due to cantori or birth of higher-order island chains. Due to this restriction, the regular parts of phase space of the sequence of 
cut-off billiards are, strictly speaking, only a fraction of the regular phase space of the original KAM billiard.

\section{Conclusions and open problems}
\label{sec:conclusions}
In this paper we demonstrated numerically that it is possible to approximate two-dimensional billiards with divided phase space by a sequence of billiards
with divided phase space and a finite number of KAM islands. Each billiard in this sequence has one more KAM island (or more precisely, island chain corresponding to a specific stable periodic orbit) than the previous billiard. Moreover each KAM island for any billiard in this sequence is a sub-island of some KAM island in the initial billiard which has, as a typical Hamiltonian system with divided phase space, an infinite number of KAM islands. 
Therefore, even though the boundary of this billiard sequence piecewise converges to the boundary of the original KAM billiard, the area of each island centered around a fixed stable orbit is, in general, decreasing along the sequence. 
Nevertheless, finite time dynamics seem to be approximated arbitrarily well with sufficiently late members of the approximating billiard sequence.
Although the consideration of billiards is easier because one can efficiently make cuts
in configuration space (billiard table), we are quite confident that this approach will prove to be efficient for approximating generic Hamiltonian systems with
divided phase space (and thus with infinite number of families of KAM-tori) by a sequences of Hamiltonian systems with sharply divided phase space with finite and increasing numbers of such KAM-tori families.

\section*{Acknowledgements}
The work has been supported by the grants P1-0044, N1-0025 of Slovenian Research Agency (ARRS), Advanced grant of European Research Council (ERC), No. 694544 - OMNES and NSF grant DMS-1600568.

\end{document}